\documentclass[aps,prb,twocolumn,floatfix,floats,superscriptaddress]{revtex4}
\usepackage{graphicx,latexsym,t1enc}

\begin{document}

\title{Net current generation in a 1D quantum ring
at zero magnetic field}

\author{Sigridur Sif Gylfadottir}
\affiliation{Science Institute, University of Iceland, 
            Dunhaga 3, IS-107 Reykjavik, Iceland}
\author{Marian Ni\c{t}\u{a}}
\affiliation{Institute of Physics and Technology of Materials,
             P.O.\ Box MG7, Bucharest-Magurele, Romania}
\author{Vidar Gudmundsson}
\affiliation{Science Institute, University of Iceland, 
        Dunhaga 3, IS-107 Reykjavik, Iceland}
\author{Andrei Manolescu}
\affiliation{Science Institute, University of Iceland, 
        Dunhaga 3, IS-107 Reykjavik, Iceland}
%

\begin{abstract} 
We study a non-adiabatic excitation of an electron system in a 1D quantum 
ring radiated by a short THz pulse. The response of two models, a continuous and discrete, 
is explored. By introducing a spatial asymmetry in the external perturbation a net current 
can be generated in the ring at a zero magnetic field. Effect of impurities 
and ratchets are investigated in combination with symmetric and asymmetric 
external excitation.
\end{abstract}


\maketitle
\section{Introduction}
Non-adiabatic generation of current in coherent quantum rings on the nanometer
scale has attracted attention 
recently.\cite{Gudmundsson03:161301,Moskalets03:075303,Moskalets03:161311}  
A one-dimensional (1D) quantum ring in its ground state exhibits a circulating 
current (persistent current) if the ring is placed in an external magnetic 
field perpendicular to the plane of the ring. The magnetic field breaks the 
left/right symmetry and the occupation of states carrying equal but opposite 
current is distorted. By subjecting the ring to a strong perturbation the 
persistent current can be changed non-adiabatically as the occupation of the 
single-electron states changes\cite{Gudmundsson03:161301,Gylfadottir04:T}.

Here we consider two models of a one-dimensional ring, continuous and discrete, and 
investigate the response to a short-lived THz perturbation. The electrons confined 
to the ring are spinless and Coulomb interactions are neglected. The main objective is 
to see whether a current can be induced in the ring at a zero magnetic field 
by introducing a left/right asymmetry in the radiation pulse. 
The comparison between the two models shows identical 
results at least for the system parameters considered in the present work.
We will also look at how impurities 
and ratchet-like potentials affect the resulting current.

\section{The Model}
\subsection{Continuous Model}
We consider a 1D quantum ring of radius $r_0$ containing a few noninteracting, 
spinless electrons. The ring is described by the following unperturbed Hamiltonian
in polar coordinates:
\begin{equation}\label{h0}
      H_0=-\frac{\hbar^2}{2m^* r_0^2}
      \frac{\partial^2}{\partial\theta^2}
\end{equation}
where $m^*$ is the effective mass of electrons.
The Hamiltonian is symmetric under any  rotation transformation. Consequently 
$H_0$ commutes with the angular momentum operator 
$L_z=-i\hbar\partial_{\theta}$.
The common set of the eigenfunctions for the two commuting operators $H_0$ and $L_z$ are 
\begin{equation}\label{esh0}
      \psi_l(\theta)
      = \frac{e^{-il\theta}}{\sqrt{2\pi r_0}}
\end{equation}
with $l=0, \pm 1, \pm 2,...$. The eigenvalue equations for the two operators are
\begin{eqnarray}\label{evh0}
      &&H_0\psi_l(\theta)=E_l\psi_l(\theta)=\frac{\hbar^2 l^2}{2m^*r_0^2}\psi_l(\theta)\\
      \label{evlz}
      &&L_z\psi_l(\theta)=-l\hbar \psi_l(\theta).
\end{eqnarray}
Except for $l=0$ all the states are doubly degenerate and carry a net persistent 
current that is normally quantified by the expectation value of the orbital 
magnetization operator
\begin{eqnarray}\label{mz}
      M_o=\frac{1}{2c}{\bf r}\times{\bf j}=-\frac{\mu_B}{\hbar}L_z.
\end{eqnarray}

\subsection{Discrete Model}
The discrete Hamiltonian of the 1D ring\cite{Aldea94:11879} is
\begin{eqnarray}\label{esh1}
      H_0=&& 2V\sum_{n}|n\rangle\langle n|\nonumber\\
      &&-V\sum_{n}\left[~ |n\rangle\langle n+1|
      +|n\rangle\langle n-1|~\right],
\end{eqnarray}
and the angular momentum operator
\begin{eqnarray}
      L_z=\frac{\hbar}{i}\frac{N}{4\pi}
      \sum_{n}\left[~ |n\rangle\langle n+1|-|n\rangle\langle n-1|~ \right].
\end{eqnarray}
The vectors $|n \rangle$ with $n=1\cdots N$ define the $N$ points of the
ring of radius $r_0$ separated by a linear distance $a=2\pi r_0/N$. 
The energy unit is $V=\hbar^2/(2ma^2)$.
The common set of eigenvectors of the two commuting operators, $H_0$ and $L_z$,
is
\begin{equation}
\label{dmeig}
     |\psi_l\rangle=\frac{1}{\sqrt{N}} \sum_n e^{-i\theta_n l}|n\rangle
\end{equation}
with $l=0,\pm1,\pm2,...,\pm(N/2-1), N/2$ (for N even)
and $\theta_n=\frac{2\pi n}{N}$ is the angle corresponding to the point $n$ of the ring
(in polar coordinates). 
The eigenvalue equations for the two operators are
\begin{eqnarray}\label{evh1}
      &&H_0|\psi_l\rangle=E_l|\phi_l\rangle=\left(2V-2V\cos ({2\pi l}/{N})\right)|\psi_l\rangle\\
      &&L_z|\psi_l\rangle=-\frac{\hbar N}{2\pi}\sin ({2\pi l}/{N})|\psi_l\rangle.
\end{eqnarray}
In the limit $l/N \to 0$ the eigenvalue set from the previous equations recover
the values from the continuous model (eqs.\ (\ref{evh0}) and (\ref{evlz})).
If the ring contains impurities or ratchets, its Hamiltonian will be completed with
the diagonal energies $\epsilon_n$
\begin{eqnarray}\label{himp}
      H_0\to H_0+\sum_n \epsilon_n |n\rangle \langle n|
\end{eqnarray}
at chosen points on the ring. We define impurities by adding a diagonal energy 
$\epsilon_n = \epsilon_{imp}$ at three selected points $n=1,20,40$.
In case of ratchets\cite{Chang02,Reimann02}, 
four equidistant saw-tooth potentials are placed at the points $n=5,\ 15,\ 25,\ 35$
of the ring. The saw-tooth potential is defined by a series of three consecutive energies:
\begin{equation}
      \epsilon_{n-1} = 0.1\epsilon_{rt}, \ \epsilon_n = 0.2\epsilon_{rt}, 
      \ \epsilon_{n+1} = 0.3\epsilon_{rt}.
\end{equation}
The net current in the discrete model is also
calculated from the expectation value of the orbital magnetization (eq.\ (\ref{mz})).

\subsection{Time dependent perturbation}
\begin{figure}
\includegraphics[width=8cm]{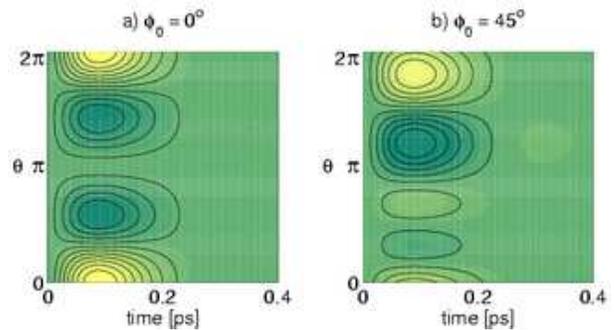}
\caption{ (Color online) Spatial distribution of the external potential pulse as a function of time 
          for (a) $\phi_0 = 0$ and (b) $\phi_0 = 45^{\circ}$. The parameter values 
          are listed in the caption of Fig.\ \ref{fig2}. Bright/yellow regions 
          indicate maxima in the potential.}
\label{fig1}
\end{figure}

At $t=0$ the quantum ring is radiated by a short THz pulse of 
duration $t_f\sim 10/\Gamma$ and frequencies $\omega_1$, $\omega_2$:
\begin{eqnarray}\label{ht}
      &&H_r(t)=V(\theta)~W(t)\\
      &&W(t)=e^{-\Gamma t}\sin (\omega_1t)\sin (\omega_2t).
\end{eqnarray}
The spatial component of the external pulse $V(\theta)$ is given by a combination of a dipole and a 
rotated quadrupole with amplitude $A$:
\begin{eqnarray}\label{vtheta}
      V(\theta)=A\left(\cos \theta+\cos 2(\theta+\phi_0)\right).
\end{eqnarray}
The deflection angle $\phi_0$ between the dipole and quadrupole fields makes 
the external perturbation asymmetric for any values $\phi_0\ne 0,\pi$. The 
perturbation is shown in Fig.\ \ref{fig1}.
To follow the time-evolution of the system we use the density matrix to specify 
the state of the system.
The ground-state density operator $\rho(t=0)$ is constructed 
in terms of a given set of the eigenvectors
$|\alpha\rangle$ of the initial time independent Hamiltonian
\begin{equation}
    \label{eq:densmat}
    \rho(t=0)=\sum_\alpha f(\epsilon_\alpha-\mu)\,|\alpha\rangle\langle\alpha|,
\end{equation}
with the Fermi distribution $f$. The equation of motion for the 
density operator 
\begin{equation}
    i\hbar\partial_t\rho(t)=[H_0+H_r(t),\rho(t)]
\end{equation}
is inconvenient for numerical evaluation so we resort instead to the time-evolution 
operator $T$, defined by $\rho(t)=T(t)\rho(0)T^{\dagger}(t)$, leading to a
simpler equation of motion 
\begin{eqnarray}\label{teo}
      i\hbar\partial_tT(t)=H(t)T(t)
\end{eqnarray}
with the initial condition $T(0)=1$. We solve the equation in the truncated 
basis of eigenvectors $\psi_l$ (eq.\ (\ref{esh0})) for the continuous Hamiltonian
(eq.\ (\ref{h0}))
or in the finite basis of eigenvectors of the discrete Hamiltonian (eqs.\ (\ref{esh1}) or (\ref{himp}))
using the Crank-Nicholson scheme.\cite{Gudmundsson03:161301}
We calculate the time evolution of the system and extend especially 
the calculation of the induced 
current in the ring to times after the dependent perturbation vanishes. 
For quantifying the currents induced by $H_r(t)$ we calculate the expectation value of 
the orbital magnetization $M_o$ (eq.\ \ref{mz}) in terms of the density matrix 
\begin{equation}\label{avo}
      {\cal{M}}_o(t)= {\rm Tr} \left[M_0\cdot \rho(t)\right].
\end{equation}
When the perturbation vanishes the time evolution of the system will be 
described by the time-evolution operator generated by the non-perturbed 
(and time independent) Hamiltonian $H_0$. 
The energy of the system will be constant
\begin{eqnarray}
      E(t>t_f)={\rm Tr} \left[H_0\cdot \rho(t)\right]={\textstyle \sum}_l~ E_l \rho_{ll}(t_f),
\end{eqnarray}
and also the magnetization for a 1D pure ring
\begin{eqnarray}\label{indm}
      {\cal{M}}_o(t>t_f)={\rm Tr} 
      \left[M_o\cdot \rho(t)\right]=-\mu_B{\textstyle \sum}_l~ l\rho_{ll}(t_f).
\end{eqnarray}
We note that the non-zero values of the induced magnetization in Eq.\,(\ref{indm})
can be obtained only by different occupation of opposite current carrying states
$\psi_l$ and $\psi_{-l}$ ($\rho_{ll}\ne \rho_{-l-l}$ at $t\simeq t_f$).
On the other hand, the magnetization for a ring with a finite width will oscillate in time 
in the presence of a perpendicular magnetic field due to coupling of radial and
angular modes of density oscillations, plasma oscillations.\cite{Gudmundsson03:161301} 

\section{Results}
\begin{figure}
\includegraphics[width=8cm]{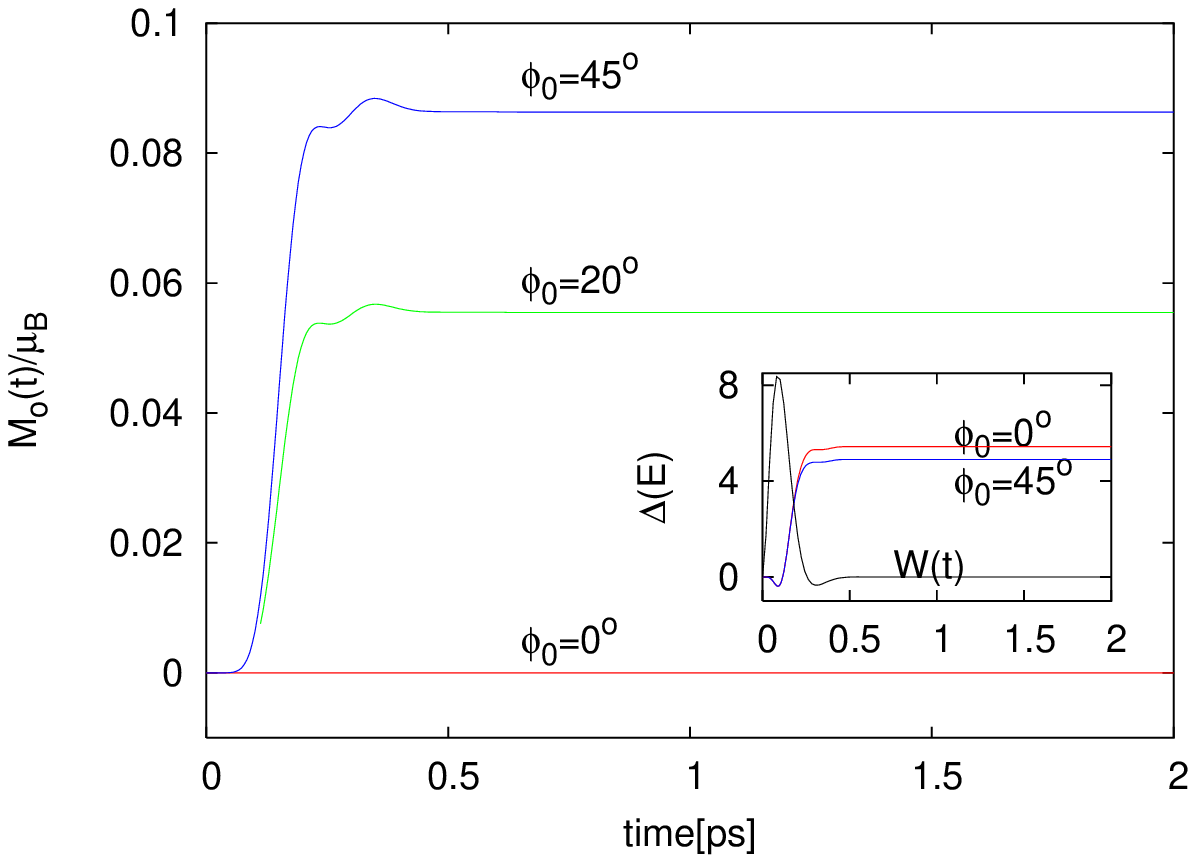}
\includegraphics[width=7.5cm]{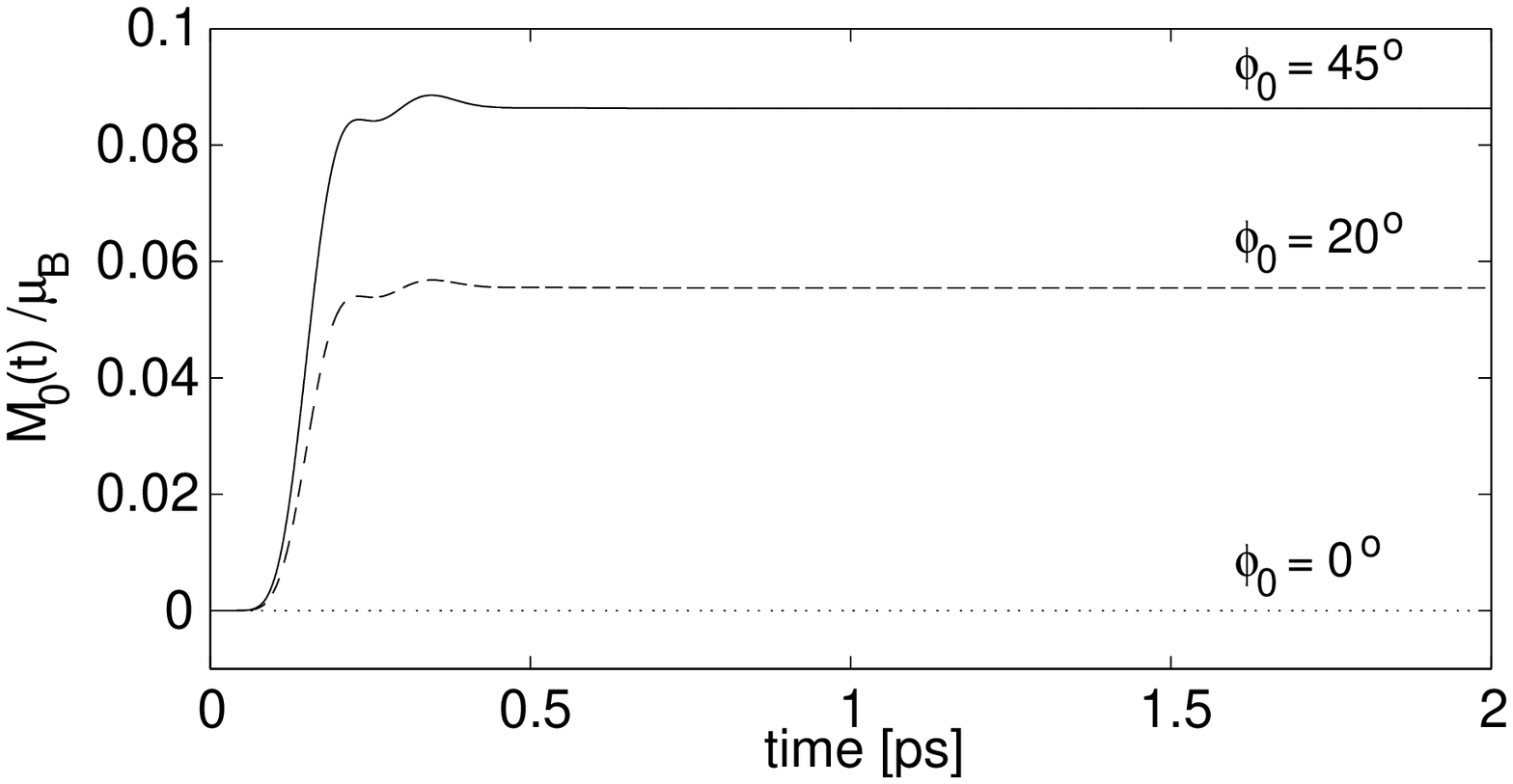}
\caption{(Color online) The time evolution of the orbital magnetization ${\cal M}_o(t)$
         calculated with the discrete model (upper) and continuous model (lower) 
         for the 1D ring described in the text. 
         Three values of the deflection angle are presented, 
         $\phi_0=0^{\circ},20^{\circ},45^{\circ}$. 
         The parameters of the external perturbation are 
         $\hbar\omega_1=2.83$meV, 
         $\hbar\omega_2=8.11$meV, 
         $\hbar\Gamma=4\hbar\omega_1=11.32$meV and
         $A=67.68$meV. 
         The frequencies are tuned to be comparable with the Bohr frequencies 
         $\hbar\omega_{0,1}=2.89$meV and $\hbar\omega_{1,2}=8.6$meV.
         The inset presents the time evolution of the absorbed energy (in meV) 
         of the discrete system for $\phi_0=0^{\circ},45^{\circ}$ and the time dependent 
         part of the external perturbation W(t).
         }
\label{fig2}
\end{figure}

We use the effective mass of an electron in a GaAs, $m^*=0.067m_e$.
For modeling the ring with $n_e=3$ electrons we set the radius to $r_0=14$nm and 
in the discrete model use $N=40$ points. Initially the system is in its ground state 
where right and left rotating states, $\left|\,l\,\right>$ and $\left|\,-l\,\right>$, 
are equally occupied and the current consequently zero (${\cal M}_o(0)=0$). In Fig.\ 
\ref{fig2} we see that if the external pulse applied to the ring, like the ring itself, 
is left/right symmetric ($\phi_0=0, \pi$) a current can not be induced since the 
occupation of states remains symmetric.
However, by introducing an asymmetry to the perturbation a constant
current along the ring appears in both the continuous and discrete models. 

\begin{figure}
\includegraphics[width=7.5cm]{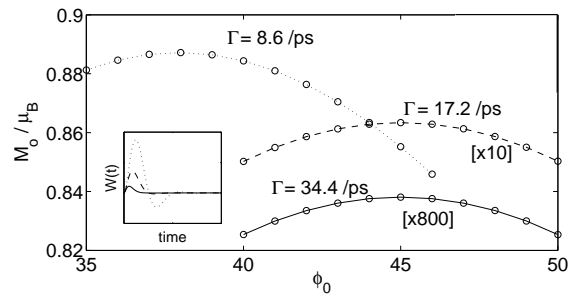}
\caption{The induced values of the orbital magnetization ${\cal M}_o$ vs.\ the 
         deflection angle $\phi_0$ for three different values of $\Gamma$ 
         (continuous model). The external perturbation parameters are the 
         same as in Fig.\ 2.
         The scale of the two lower curves has been magnified as indicated. 
         The inset depicts the time dependent part of the external perturbation 
         for the three situations.
        }
\label{fig4}
\end{figure}

For a weak excitation a perturbation analysis to 2nd order (in no external 
magnetic field) reveals that,
for a single electron initially in the ground state
$\left|\,0\,\right>$,
the difference between the transition probabilities to the states $\left|\,1\,\right>$ and 
$\left|\,-1\,\right>$
is proportional to $\sin2\phi_0$
\begin{equation}\label{probsin}
{\cal{P}}_{0,1}(t)-{\cal{P}}_{0,-1}(t) = F(t)\,\sin2\phi_0
\end{equation}
where $F(t)$ is a double integral depending on the parameters of the external 
perturbation. This implies that for a deflection angle of $\phi_0=45^{\circ}$ 
the resulting current is at maximum. Fig.\ \ref{fig4} shows the induced  
magnetization (continuous model) after the perturbation vanishes ($t>t_f$) as 
a function of the deflection  angle $\phi_0$ for three values of $\Gamma$. 
\begin{figure}
\includegraphics[width=7.2cm]{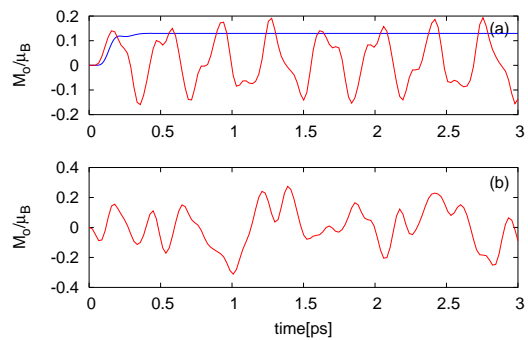}
\caption{ (Color online) (a) The orbital magnetization ${\cal{M}}_o(t)$ as a function of time
           calculated with the discrete model of the 1D ring.
           The external perturbation is asymmetric ($\phi_0=45^{\circ}$)
           with the same parameters as in Fig.\ 2 (except for $A$=135meV).
           The ring contains four ratchets with $\epsilon_{rt}=V$
           (explained in text). The almost constant blue line is
           the time evolution of ${\cal{M}}_o(t)$ for a pure ring.
           (b) The ring with ratchets plus impurities present ($\epsilon_{imp}=0.2V$).
           $V=\hbar^2/2m^*a^2=117.5$meV.
           }
\label{fig5}
\end{figure}
Clearly for strong excitation (lower $\Gamma$ in Fig.\ \ref{fig4}) the results do 
not depend on $\phi_0$ as eq.\ (\ref{probsin}) shows, but have a more complex dependence on the exact 
form of the perturbation. As $\Gamma$ 
decreases the maximum of the current shifts away from $\phi_0=45^{\circ}$. The identical 
behavior is seen in the discrete model. 

\begin{figure}
\includegraphics[width=6.8cm]{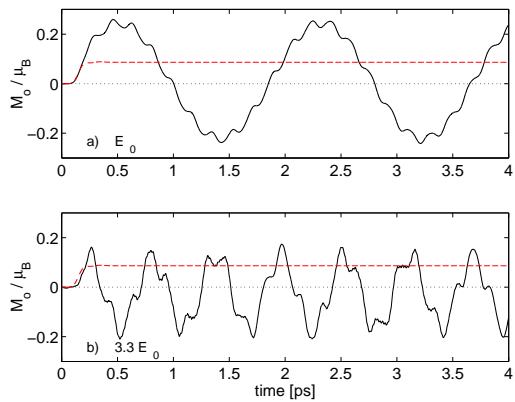}
\caption{  (Color online) The orbital magnetization ${\cal{M}}_o(t)$ as a function of time calculated with
           the continuous model of the 1D ring. 
           The external perturbation is asymmetric ($\phi_0=45^{\circ}$) 
           with the same parameters as in Fig.\ 2. The ring contains four ratchets 
           (explained in text). The strength of ratchet potential $\epsilon_{rt} = E_0$ in (a)
           and $\epsilon_{rt}=3.33E_0$ in (b). The dashed red curve is the time 
           evolution of ${\cal{M}}_o(t)$ for a pure ring. 
           $E_0=\hbar^2/2m^*r_0^2=2.90$meV.
           }
\label{fig6}
\end{figure}

The constant magnitude of the current generated in the system by a left/right asymmetric
excitation can be destroyed by the inclusion of ratchets and impurity potential 
as can be seen in Fig.\ \ref{fig5}. 
In Fig.\ \ref{fig6} we see that the frequency 
of current oscillations depends heavily upon the strength of the ratchet potential. 
As the height of the ratchets increases the frequency grows. Furthermore the 
reduction of the current occurs even though the ratchet potential is very weak. It 
is interesting that the period of oscillation can be much larger than that of the applied 
perturbation. 
\begin{figure}
\includegraphics[width=7.2cm]{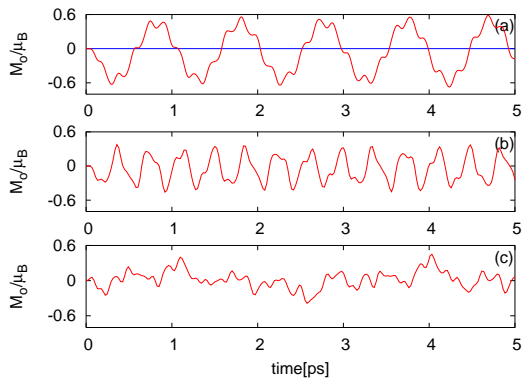}
\caption{(Color online) Oscillations of the orbital magnetization ${\cal{M}}_o(t)$ calculated with the discrete
         model. 
         The ring is radiated with a symmetric pulse ($\phi_0=0^{\circ}$). The perturbation 
         parameters are the same as in Fig.\ 2 (except for $A$=135meV).
         (a) Four sticks symmetrically placed on the ring. Each stick is defined by adding a 
         diagonal energy $\epsilon_{n}=0.2V$ at points $n=5,15,25,35$ of the ring. 
         The constant line is the time evolution of ${\cal{M}}_o(t)$ for a pure ring.
         (b) Four ratchets on the ring ($\epsilon_{rt}=V$). 
         (c) Same as (b) plus impurities present ($\epsilon_{imp}=0.2V$). 
         $V=\hbar^2/2m^*a^2$.
           }
\label{fig7}
\end{figure} 
The reader may now wonder if it would be possible to excite a current in the
system by applying a left/right symmetric external electric field, but at the
same time having an asymmetric static potential on the ring. We have tried
several asymmetric potentials on the ring and different strength of the
external excitation, but have not been successful in generating a steady
DC current. Some results are seen in Fig.\ \ref{fig7}. Clearly, as we have
drawn attention to above, the frequency of the current oscillations can be made
very low compared to the dominant frequencies included in the excitation pulse.

\section{Conclusions}
For a quantum ring in no external magnetic field a left/right asymmetric excitation
is essential for the appearance of a circulating current. In an external magnetic 
field the energy of states whose orbital angular momentum is aligned with the field
is lowered, thus breaking the symmetry and generating a DC current.
 Here we have seen that a similar effect 
can be accomplished by an asymmetric excitation pulse at zero magnetic field, but 
a DC current can not be generated by a left/right symmetric excitation even though 
the ring itself is made asymmetric by a static potential or by change to the geometry,
breaking the perfect circular symmetry. The introduction of a ratchet potential
in the ring that is excited by an asymmetric external field changes the DC current
created into an AC current with arbitrary low frequency depending on the intensity
of the ratchet. This is a clear nonlinear behavior that could not be observed 
by traditional linear response methods.  

The electron-electron interaction has been neglected here as the effects studied
do not depend on it qualitatively, but certainly quantitatively they do.\cite{Chakraborty94:8460}

\begin{acknowledgments}
      The research was partly funded by the Icelandic Natural Science Foundation,
      and the University of Iceland Research Fund.
      M.N. was supported by a NATO Science Fellowship and Ceres.
      We acknowledge instructive discussion with Dr.\ Cheng-Hung Chang,
      and prof.\ A.\ Aldea.
\end{acknowledgments}

%
%
\bibliographystyle{apsrev}

%
%
%
\end{document}